\documentclass[12pt]{iopart}
\usepackage{graphicx,amsfonts,amsbsy, amssymb, amsthm}

\newcommand{\R}{{\mathbb R}}

\newcommand{\Z}{{\mathbb Z}}

\newcommand{\coloneq}{\mathbin{\hbox{\raise0.08ex\hbox{\rm :}}\!\!=}}
\newcommand{\eqcolon}{\mathbin{=\!\!\hbox{\raise0.08ex\hbox{\rm :}}}}

\renewcommand{\leq}{\leqslant}
\renewcommand{\geq}{\geqslant}
\renewcommand{\epsilon}{\varepsilon}

\newcommand{\dpdot}{{\lower0.33ex\hbox{\LARGE$\cdot$}}}
\newcommand{\edpdot}{{\lower0.33ex\hbox{\Large$\cdot$}}}
\renewcommand{\le}{\leqslant}
\renewcommand{\ge}{\geqslant}

\newtheorem{theorem}{Theorem}[section]
\newtheorem{proposition}[theorem]{Proposition}
\newtheorem{definition}[theorem]{Definition}

\newtheorem{lemma}[theorem]{Lemma}

\newtheorem{conjecture}[theorem]{Conjecture}
\newcommand{\dimostrazione}{\noindent{\sl Proof.}\phantom{X}}
\renewcommand{\theequation}{\thesection.\arabic{equation}}

\newcommand{\eqref}[1]{\eref{#1}}

\newcommand{\aut}{\mbox{\rm Aut}}

\def\CC{\Bbb{C}}
\def\RR{\Bbb{R}}
\def\Z{\Bbb{Z}}
\def\ZZ{\Bbb{Z}}
\def\TT{\Bbb{T}}

\newcommand{\beq}{\begin{equation}}
\newcommand{\eeq}{\end{equation}}
\newcommand{\beqa}{\begin{eqnarray}}
\newcommand{\eeqa}{\end{eqnarray}}

\newcommand{\bdm}{\begin{displaymath}}
\newcommand{\edm}{\end{displaymath}}

\newcommand{\ud}{\mathrm{d}}

\begin{document}

\title[Spectral edges of periodic operators]
{On occurrence of spectral edges for periodic operators inside the Brillouin zone}
\author{J.~M.~Harrison$^1$, P.~Kuchment$^1$, A.~Sobolev$^2$ and B.~Winn$^1$}
\address{$^1$ Mathematics Department, Texas A\&M University,
College Station, Texas 77843-3368, USA}
\address{$^2$ School of Mathematical Sciences,  University of Birmingham,
Edgbaston, Birmingham, B15 2TT, UK}
\begin{abstract}
The article discusses the following frequently arising question on
the spectral structure of periodic operators of mathematical
physics (e.g., Schr\"{o}dinger, Maxwell, waveguide operators,
etc.). Is it true that one can obtain the correct spectrum by
using the values of the quasimomentum running over the boundary of
the (reduced) Brillouin zone only, rather than the whole zone? Or,
do the edges of the spectrum occur necessarily at the set of
``corner'' high symmetry points? This is known to be true in $1D$,
while no apparent reasons exist for this to be happening in higher
dimensions. In many practical cases, though, this appears to be
correct, which sometimes leads to the claims that this is always
true. There seems to be no definite answer in the literature, and
one encounters different opinions about this problem in the
community.

In this paper, starting with simple discrete graph operators, we
construct a variety of convincing multiply-periodic examples
showing that the spectral edges might occur deeply inside the
Brillouin zone. On the other hand, it is also shown that in a
``generic'' case, the situation of spectral edges appearing at
high symmetry points is stable under small perturbations. This
explains to some degree why in many (maybe even most) practical
cases the statement still holds.
%
%
\end{abstract}
\maketitle
\section{Introduction}\label{S:intro}
The article discusses the following frequently arising question on
the spectral structure of periodic operators of mathematical
physics, which in particular is prominent due to the recent surge
in studying photonic crystals \cite{FK1}-\cite{FK4},
\cite{Ku_CONM}-\cite{Ku_thin}. Let us have a periodic elliptic
self-adjoint operator $L(x,D)$ (e.g., Schr\"{o}dinger, Maxwell),
where we use the standard notation $D=\frac{1}{\rmi}\nabla$. The
operator is considered in the whole space $\RR^n$, or in a
periodic domain (on a periodic periodic manifold), e.g. in a
periodic waveguide. The standard Floquet-Bloch theory (e.g.,
\cite{AM,Ku,RS}) shows that the spectrum of $L$ in the infinite
periodic medium can be obtained as follows: one fixes a value $k$
of the quasimomentum in the first Brillouin zone $B$, finds the
(discrete) spectrum of the corresponding Bloch Hamiltonian
$L(k)=L(x,D+k)$ acting on periodic functions, and then takes the
union over all quasimomenta in the Brillouin zone. The question we
address in this work is whether the correct spectrum can be
obtained as the union over the boundary of the Brillouin zone 
only\footnote{If
additional symmetries are present in the system, one considers the
reduced (with respect to these symmetries) Brillouin zone. 
Another version of this question is
whether the edges of the spectrum are attained on the set of high symmetry points
of the Brillouin zone only. The importance
of such points has been known since at least the paper
\cite{BouSmolWigner}. When one needs to find the 
density of states, the full Brillouin
zone is always required.}
%

This is well known to be true in $1D$
(e.g., \cite{Ea,RS}). In particular, the edges of the spectrum
occur at the spectra of the periodic and anti-periodic problems on
the single period. If this claim is correct in higher dimensions,
the computational task is significantly simplified, due to reduced
dimension. This is important, for instance, in optimization
procedures, when one needs to run the spectral computation at each
iteration \cite{CD,CD2}. An experimental observation is that in
most practical cases this is correct. One frequently encounters
the belief that this is always true (albeit no justification is
ever provided). On the other hand, unlike in $1D$, there is no
analytic reason for this property to hold. Moreover, many
researchers are aware that numerics sometimes produces
counterexamples. Surprisingly, such examples are hard to come by
and are usually not very convincing for an analyst (e.g., the
error in computing the spectrum using only the boundary of the
Brillouin zone is usually very small). The experience is that one
needs to make the medium inside the fundamental domain
(Wiegner-Seitz cell) truly asymmetric to achieve such examples.

The first goal of this text is to provide simple definite examples to
disprove the claim that the edges of the spectral bands can be
found by using the boundary of the (reduced) Brillouin zone only. This is
done by first analyzing some
discrete graph systems. Section \ref{S:graph}
describes such combinatorial graph counterexamples. Section
\ref{S:quantum} deduces from this some quantum graph (see \cite{Ku_graphs})
examples. Then in Section \ref{S:guide}, we bootstrap this to examples of
waveguide systems or Laplace-Beltrami operators on thin tubular
branching manifolds. Possibilities for obtaining counterexamples
of the Schr\"{o}dinger and Maxwell cases are discussed in Section
\ref{S:Schroed}.

It is surprising, however, that the claim that we show to be
incorrect in general, is still correct (or almost correct) so
often. Thus, in many practical cases, computations along the
boundary of the (reduced) Brillouin zone (and as a matter of fact,
often at high symmetry corner points only) provide the correct spectrum.
We suggest
an explanation of this effect in the final  Section \ref{S:positive}. There we
attempt to explain how
it can happen that one often sees the spectral edges occurring at
the high symmetry points only. It is shown that ``generically''
this occurrence is stable under small perturbations. In other
words, there are open sets of ``good'' and ``bad'' periodic
operators, the boundary between which consists of non-generic
operators.  This
probably explains the frequent occurrence of the effect in practice.

Finally, the last sections provide additional remarks
and acknowledgments.

\section{Combinatorial graph examples}\label{S:graph}

We start by considering difference operators acting on a periodic
graph. These will serve to illustrate the general ideas in a
situation which is not difficult to analyze. Furthermore, building
upon them, we will provide examples of more complex periodic
spectral problems with the desired spectral feature.

\subsection{The main graph operators}\label{SS:graphs}
We consider the $\Z^2$-periodic planar graph $\Gamma$, with the
fundamental domain $W$ shown in Figure \ref{fig:quantumfd} below.
\begin{figure}[ht]
\begin{center}
\setlength{\unitlength}{4cm}
\begin{picture}(1,1)
\put(0.0,0.0){\includegraphics[width=4cm]{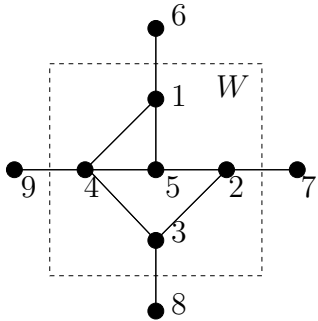}}
\put(0.26,0.41){$4$} \put(0.53,0.41){$5$} \put(0.74,0.41){$2$}
\put(0.55,0.257){$3$} \put(0.55,0.71){$1$} \put(0.7,0.74){$W$}
\put(0.05,0.41){$9$} \put(0.55,0.02){$8$} \put(0.98,0.41){$7$}
\put(0.55,0.98){$6$}
\end{picture}
\caption{The graph $\Gamma$ with fundamental region $W$.}
\label{fig:quantumfd}
\end{center}
\end{figure}
One imagines the graph $\Gamma$ as obtained by tiling the plane
with the $\Z^2$-shifted copies of $W$. We will label the vertices
in $W$ and near $W$ with the numbers shown in Figure
\ref{fig:quantumfd}.

Let $\ell^2(\Gamma)$ be the Hilbert space of square-summable
functions defined on the set of vertices of $\Gamma$. The discrete
Laplacian on $\Gamma$ can be defined in several (not always
equivalent) ways (e.g., \cite{Chung,Colin}). We will use two of
these.

The first one, $\Delta$, is defined for $f\in\ell^2(\Gamma)$ by
\begin{equation} \label{eq:uno}
  (\Delta f)(v) \coloneq \sum_{u\sim v} f(u),
\end{equation}
where the notation $u\sim v$ means that vertex $u$ is adjacent
(connected by an edge) to vertex $v$. For instance,
\begin{equation*}
    (\Delta f)(5)=f(1)+f(2)+f(4).
\end{equation*}
The Laplacian defined in \eqref{eq:uno} differs from another
discrete Laplacian often used in the literature by the term $d_v
f(v)$, where $d_v$ is the degree (valence) of the vertex $v$:
$\sum_{u\sim v} f(u)-d_v f(v)$.

We will also employ another version, $L$, of the Laplacian, which
is defined as
\begin{equation}
  \label{eq:due}
  (Lf)(v) \coloneq \frac{1}{\sqrt{d_v}}\sum_{u\sim
  v}\frac1{\sqrt{d_u}}f(u).
\end{equation}
One could call it the Laplace-Beltrami operator. The need for this
operator in our study will become clear when we move to the
quantum graph case.

One notices that both $\Delta$ and $L$ are bounded operators in
$\ell^2(\Gamma)$.

In the case of a regular graph (i.e. the one with constant degrees
$d_v=d$ of vertices), the spectra of $\Delta$ and $L$ can be
easily related. However, our graph $\Gamma$ is not regular, and
thus these spectra need to be studied independently.

The following statement is well known (e.g., \cite{Chung}) and
immediate:
\begin{lemma}\label{L:autom}
The operators $\Delta$ and $L$ commute with any automorphism
$T\in\aut(\Gamma)$ of the graph $\Gamma$. In particular, they
commute with $\ZZ^2$-shifts on $\Gamma$.
\end{lemma}
For $p=(p_1,p_2)\in\ZZ^2$, we denote by $T(p)$ the shift
operator by $p$ on $\Gamma$. E.g., $T(1,0)$ shifts the vertex $4$
to the vertex $7$ and $T(0,1)$ shifts $3$ to $6$.

Due to the periodicity of the operators, one can use the standard
Floquet-Bloch theory \cite{AM,Ea,Ku,RS} to study their spectra. In
the particular case of graphs, this theory is also described, for
instance, in \cite{EG}, \cite{Ku_difference}-\cite{Ku_graphs2},
\cite{Oleinik}.

Let $k=(k_1,k_2)$ be a \textit{quasimomentum} in the
\textit{Brillouin zone} $B=[-\pi,\pi]^2$. Consider the space
$\ell^2_k$ of all functions $f$ satisfying the following
\textit{Floquet (Bloch, cyclic) condition}:
\begin{equation}\label{E:cyclic}
f(T(p)v)=\rme^{\rmi p\cdot k}f(v)
\end{equation}
for all $p\in \ZZ^2$. Here $p\cdot k=p_1 k_1+p_2
k_2$. Such a function $f$ is clearly uniquely determined by the
vector $(f_1,f_2,..., f_5)^t$ of its values at the five vertices
in $W$, and thus $\ell^2_k$ is five-dimensional and naturally
isomorphic to $\ell^2(W)$.

\begin{definition}
We will denote by $\Xi$ the boundary $\partial B$ of the Brillouin
zone $B=[-\pi,\pi]^2$ and by $X$ the set of points $k=(k_1,k_2)\in
B$ such that $k_1$ and $k_2$ are integer multiples of $\pi$.
\end{definition}

We now define the Floquet Laplacian $\Delta(k):\ell^2(W)\to
\ell^2(W)$ as the restriction to the space $\ell^2_k$ of the
operator $\Delta$ defined as in (\ref{eq:uno}). In terms of the
basis of the delta functions at vertices of $W$, this operator has
the following matrix:
\begin{equation}
  \label{eq:cinque}
  \Delta(k)\coloneq  \left(
\begin{array}{ccccc}
0 & 0 & \rme^{\rmi k_1} & 1 & 1 \\
0 & 0 & 1 & \rme^{\rmi k_2} & 1 \\
\rme^{-\rmi k_1} & 1 & 0 & 1 & 0 \\
1 & \rme^{-\rmi k_2} & 1 & 0 &1 \\
1  & 1 & 0 & 1 & 0
\end{array}\right).
\end{equation}

In a similar way, one can define $L(k)$ and observe that
\begin{equation}
  \label{eq:sei}
  L(k)= S^{-1}\!\Delta(k)S^{-1},
\end{equation}
where $S$ is the matrix
\begin{equation}
  S\coloneq \left(
\begin{array}{ccccc}
\sqrt{3} & 0 & 0 & 0 & 0 \\
0 & \sqrt{3} & 0 & 0 & 0 \\
0 & 0 & \sqrt{3} & 0 & 0 \\
0 & 0 & 0 & 2 & 0 \\
0 & 0 & 0 & 0 & \sqrt{3} \\
\end{array}
\right).
\end{equation}

We now state the standard conclusion of the Floquet theory about
the relation between the spectra of $\Delta$ and $\Delta(k)$, or
$L$ and $L(k)$ (e.g., \cite{Ea, EG, Ku_difference, Ku, Ku_graphs2,
Oleinik, RS}). For each fixed $k$, the matrix $\Delta(k)$
(correspondingly, $L(k)$) is self-adjoint, and thus admits a
spectrum of $5$ eigenvalues $\{\lambda_j( k )\}_{j=1}^5$
(correspondingly, $\{\mu_j( k )\}_{j=1}^5$), which we number in
non-decreasing order. It is well known that then each of the
functions $\lambda_j( k )$ is continuous. The multiple-valued
function $ k \mapsto \{\lambda_j ( k )\}$ is called the
\textit{dispersion relation}. Its graph is the \textit{dispersion
curve}, also called the \textit{Bloch variety}. Each of the
individual functions $ k \mapsto \lambda_j ( k )$ is called the
$j$th \textit{branch} of the dispersion relation.
\begin{proposition}\cite{Ea,Ku,Ku_difference,Ku_graphs2,RS} \label{prop:uno}
The spectrum of $\Delta$ (correspondingly, $L$) is the union over
$ k \in B$ of the spectra of $\Delta ( k )$
(correspondingly, $L( k )$):
\begin{equation}
  \eqalign{
  \sigma(\Delta)=\bigcup\limits_{ k \in B} \sigma(\Delta ( k ))
  ={\bigcup\limits_{ k \in B}\bigcup\limits_{j=1}^5 \lambda_j( k )},\\
\sigma(L)=\bigcup\limits_{ k \in B} \sigma(L ( k ))
  ={\bigcup\limits_{ k \in B}\bigcup\limits_{j=1}^5 \mu_j( k )}.}  
\label{eq:sette}
\end{equation}
The segments $I_j=\bigcup\limits_{ k \in B}
\lambda_j( k )$ (and their analogs for the operator $L$) are
called {\em bands} of the spectrum of $\Delta$ (correspondingly,
of $L$).
\end{proposition}

Notice that our graph $\Gamma$ does not have any point symmetries,
and thus the reduced Brillouin zone is equal to the full one. So,
the question we would like to address is whether one can replace
the union over $ k \in B$ in (\ref{eq:sette}) by the union
along the boundary $\Xi=\partial B$ of the Brillouin zone $B$  only.
A more restricted question is whether the band edges are attained at points of
$X$ only. As we will show in the next sub-section,
both of these properties do not hold, and calculations along the boundary lead to
significant errors in spectra of $\Delta$ and $L$.

\subsection{Spectral edges - counterexamples}\label{SS:graph_edges}

In this sub-section we show that computations along $\Xi=\partial
B$ (and thus over $X$ as well) do not necessarily lead to the
correct spectra of $\Delta$ and $L$, and the errors can be
significant. We are interested in whether the segments
$I_j=\bigcup_{ k \in B} \lambda_j( k )$ and $I^\prime_j=\bigcup_{
k \in \partial B} \lambda_j( k )$ coincide. We will show that for
our examples, even the unions
$\sigma(\Delta)=\bigcup\limits_{j=1}^5 I_j$ and
$\bigcup\limits_{j=1}^5 I^\prime_j$ are different (the situation
is analogous for $L$). This means that using only the boundary of
the Brillouin zone, one does not recover the spectral edges (and
thus the spectrum) correctly.

\subsubsection{Operator $\Delta$}
One can easily find the spectrum of the Floquet Laplacian $\Delta(
k )$ (which is a simple $5\times 5$ matrix), using for instance
Matlab. Computing it for a grid in the Brillouin zone (we have
used the uniform $64\times 64$ grid in $B$), one can obtain the
whole spectrum of $\Delta$. This leads to the following numerical
values of the five bands: $[-2.73,-1.90]$,  $[-1.63,-1.00]$,
$[-0.73,0.73]$,  $[0,1.46]$, and  $[2.00,3.23]$. One notices that
there are spectral gaps present between all consecutive bands,
except the 3rd and 4th ones, which overlap.

Since it is sufficient for our purpose to provide a single
counterexample, we will focus on the second band $[-1.63,-1.00]$
only.

\begin{figure}[ht]
\begin{center}
\setlength{\unitlength}{4cm}
\begin{picture}(2,1.2)
\put(0.05,0.05){\includegraphics[angle=0,width=4cm,height=4cm]{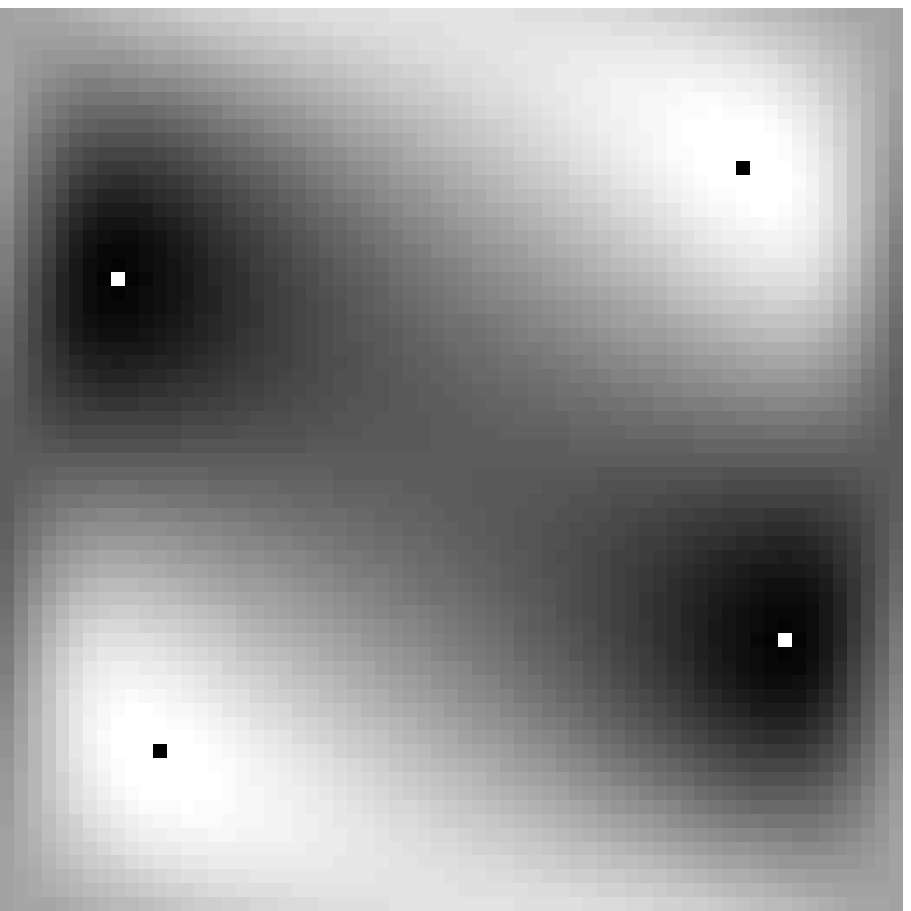}}
\put(0,0){\line(1,0){1.1}} \put(1.1,0){\line(0,1){1.1}}
\put(1.1,1.1){\line(-1,0){1.1}} \put(0,1.1){\line(0,-1){1.1}}
\put(1.4,0.05){\includegraphics[angle=0,width=1cm,height=4cm]{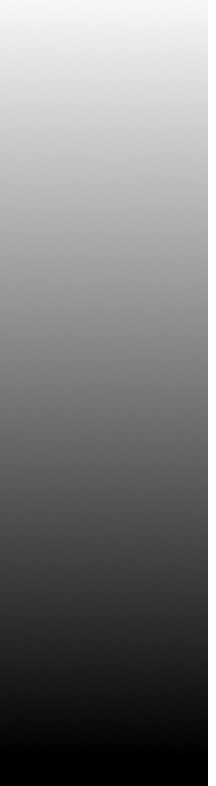}}
\put(1.35,0.05){\line(1,0){0.4}} \put(1.35,1.055){\line(1,0){0.4}}
\put(1.35,0.05){\line(0,1){1.005}}
\put(1.7,0.05){\line(0,1){1.005}} \put(1.24,1.06){$\lambda$}
\put(1.78,1.02){$-1.001$} \put(1.78,0.05){$-1.630$}
\end{picture}
\caption{A gray scale image of the second branch of the spectrum
of $\Delta ( k )$. Extrema points are highlighted.}
\label{fig:B}
\end{center}
\end{figure}

\begin{figure}[ht]
\begin{center}
\setlength{\unitlength}{5cm}
\begin{picture}(2.5,2.0)
\put(0.5,0.0){\includegraphics[angle=0,width=10cm,height=10cm]{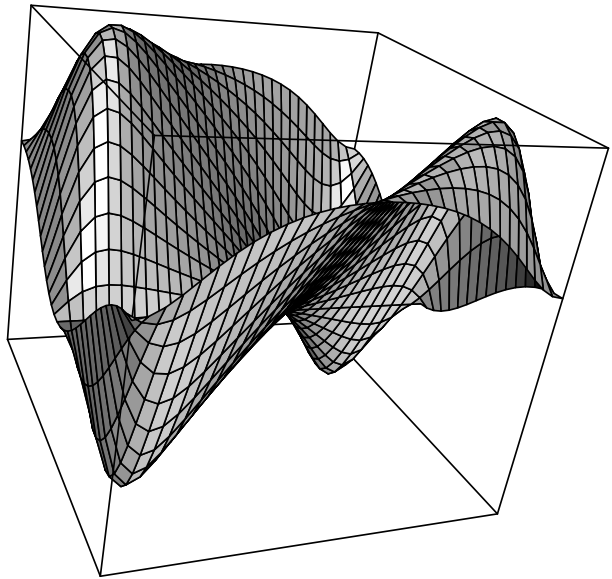}}
\end{picture}
\caption{A graph of the second branch for the graph $\Gamma$.}
\label{fig:sei}
\end{center}
\end{figure}

A gray scale plot of the second branch (corresponding to the
second band of the spectrum) is given in Figure \ref{fig:B}
\footnote{This and other plots are drawn over the square
$[0,2\pi]^2$, rather than the Brillouin zone $B=[-\pi,\pi]^2$. The
origin $(0,0)$ is located in the upper left corner.}.
This numerical evidence shows that the band edges (i.e., the
extrema of the branch function) occur at some values $ k $ not in
$\Xi$. This is confirmed by the graph of the branch presented in
Figure \ref{fig:sei}.
Let us now make this observation rigorous by finding the maximum
and minimum values of $\lambda$ on $\Xi$. The characteristic
polynomial of $\Delta ( k )$ is
\begin{eqnarray}
\fl 
c_\Delta(\lambda; k )=\lambda^5- 8\lambda^3 -(2\cos k_1 + 
4\cos k_2 +2)\lambda^2 \nonumber \\
+(8-2\cos(k_1+k_2)-4\cos k_1 -2\cos k_2)\lambda \nonumber \\
-2\cos(k_1+k_2)-2\cos(k_1-k_2)+4\cos k_2. \label{eq:char:poly}
\end{eqnarray}
Since the second band does not intersect any others, standard
perturbation theory \cite{Kato} implies that the corresponding
eigenvalue branch $\lambda( k )=\lambda_2( k )$ is analytic. It is
not hard to check all possible extremal values of $\lambda$ on
$\Xi$. Indeed, assuming that $k_1$ equals $\pm \pi$ or $0$, one
can differentiate the secular equation $c_\Delta(\lambda; k )=0$
with respect to $k_2$ and use that at an extremal point (unless it
is one of the points of $X$), one has $\frac{\partial
\lambda}{\partial k_2}=0$. One can do the same with the roles of
$k_1$ and $k_2$ reversed. This calculation shows that extremal
points can be located only where $k_1$ and $k_2$ are integer
multiples of $\pi$, i.e. at $X$. All points of $X$ can be checked
to yield that the minimum value on $X$ is $-\sqrt{2}\approx
-1.414$, attained at $ k =(\pi,0)$ and $ k =(\pi,\pi)$, and the
maximum value is $-\sqrt{4-\sqrt{8}}\approx -1.082$ at $ k
=(0,\pi)$. These values can be compared with the numerically found
extreme values of $-1.630$ at $ k \approx(1.865,0.785)$ and
$-1.001$ at $ k \approx(-1.080,5.203)$. This gives a difference of
about $8\%$ at the upper edge, and $15\%$ at the lower edge. These
values and their symmetric counterparts are highlighted in Figure
\ref{fig:B}.

In fact, the location and value of the maximum of the second
branch can be found exactly. It is not hard to check that the
value $\lambda=-1$ attained at $\displaystyle
 k _*=\left(\frac{\pi}3, \frac{5\pi}3\right)$ is a maximum.

Thus, we have an example of the situation when restricting the
search of the edges of the spectrum to quasimomenta from $\Xi$,
leads to significant errors.

One can draw other branches of the dispersion relation. They show
that some of the branches do attain their extrema on $X$ only,
while some others do not.

\subsubsection{Operator $L$}
\begin{figure}[ht]
\begin{center}
\setlength{\unitlength}{4cm}
\begin{picture}(2,1.2)
\put(0.05,0.05){\includegraphics[angle=0,width=4cm,height=4cm]{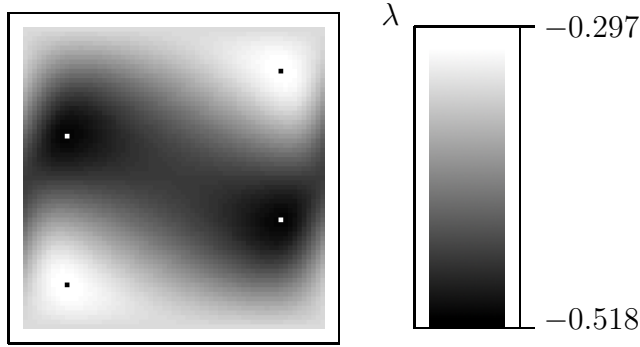}}
\put(0,0){\line(1,0){1.1}} \put(1.1,0){\line(0,1){1.1}}
\put(1.1,1.1){\line(-1,0){1.1}} \put(0,1.1){\line(0,-1){1.1}}
\put(1.4,0.05){\includegraphics[angle=0,width=1cm,height=4cm]{greyscale.eps}}
\put(1.35,0.05){\line(1,0){0.4}} \put(1.35,1.055){\line(1,0){0.4}}
\put(1.35,0.05){\line(0,1){1.005}}
\put(1.7,0.05){\line(0,1){1.005}} \put(1.24,1.06){$\lambda$}
\put(1.78,1.02){$-0.297$} \put(1.78,0.05){$-0.518$}
\end{picture}
\caption{A gray scale image of the second branch for the operator
$L$. Extrema are highlighted.} \label{fig:other:B}
\end{center}
\end{figure}

The operator $L$ can be analyzed similarly. We briefly summarize
the findings. The characteristic polynomial for $L( k )$ is
\begin{eqnarray}
\fl
c_L(\lambda; k )={\lambda}^{5}-{\frac {7}{9}}{\lambda}^{3}-
\left( \frac1{18} +\frac19 \cos
k_2  +\frac1{18}\cos  k_1  \right) {\lambda}^{2} \nonumber \\
+ \left({\frac {13}{162}} -{\frac
{7}{162}}\cos k_1 - {\frac {1}{54}}\cos  k_2  -{\frac {1}{54}}\cos
 ( k_1+k_2 ) \right) \lambda \nonumber \\ 
+{ \frac {1}{81}}\cos  k_2  -{\frac{1}{162}}\cos
 ( k_1-k_2 ) -{\frac {1}{162}}\cos ( {k_1}+k_2 ).  \label{eq:other:poly}
\end{eqnarray}
We observe that \eqref{eq:other:poly} is quite similar to
\eqref{eq:char:poly}, modulo changes in some constants. The
structure of the branches of solutions to $c_L(\lambda; k )=0$ are
also qualitatively similar to the ones for the spectrum of
$\Delta$. The spectrum of $L$ on $\Gamma$ consists of five bands:
$[-0.830,-0.606]$, $[-0.518,-0.297]$, $[-0.219,0.219]$,
$[0,0.485]$, $[0.611,1.000]$, with the third and fourth bands
overlapping. The second branch is similar in appearance to the one
of the operator $\Delta$ (see Figure \ref{fig:other:B}). It shares
the property that the band edges occur away from the set $\Xi$.

Analogously to the case of the operator $\Delta$, we again find
that the extremal values on $\Xi$ of the second band function can
only occur at the points $ k \in X$. It turns out that the
maximum value on $\Xi$ is $-1/3\approx-0.333$ at $ k =(0,0)$
and $ k =(0,\pi)$, and the minimum value is
$-\sqrt{2}/3\approx-0.471$ at $ k =(\pi,0)$ and
$ k =(\pi,\pi)$. These can be compared with the numerically
found maximum over all of $B$ of approximately $-0.297$ at
$ k \approx(5.40, 0.88)$ and $-0.518$ at
$ k \approx(0.26,0.88)$. The difference is $10.8\%$ at the
upper edge, and $9.5\%$ at the lower edge.

In summary, we have described two difference operators $\Delta$
and $L$ acting on a periodic graph $\Gamma$, which have spectra
with band edges occurring away from the boundary of the Brillouin
zone.

\subsection{An example in the presence of point
symmetries}\label{SS:symmetric} The previously described examples
dealt with a graph $\Gamma$ that had only translation invariance
with respect to $\Z^2$, and no point symmetries (i.e., symmetries
that would fix a point on the graph). Thus, the Brillouin zone was
not reduced. In this section, we provide an example where the
point symmetry group is non-trivial, while the effect we observed
in the previous sections still holds. We will also observe in some
cases that spectral edges can occur on $\Xi$, but not on $X$.

\begin{figure}[ht]
\begin{center}
\setlength{\unitlength}{4cm}
\begin{picture}(2,1.9)
\put(0.0,0.0){\includegraphics[angle=0,width=8cm,height=7cm]{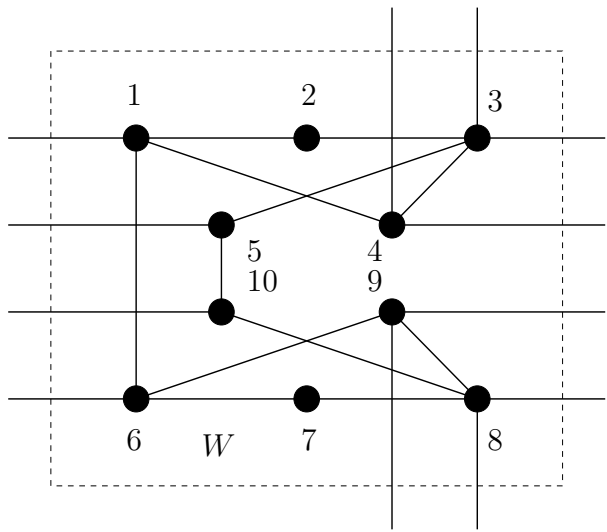}}
\put(1.2,0.9){$4$} \put(0.8,0.9){$5$} \put(0.98,1.42){$2$}
\put(0.65,0.25){$W$} \put(0.4,1.42){$1$} \put(1.6,1.40){$3$}
\put(0.4,0.27){$6$} \put(0.98,0.27){$7$} \put(1.6,0.27){$8$}
\put(1.2,0.8){$9$} \put(0.8,0.8){$10$}
\end{picture}
\caption{The fundamental domain $V$ of the $\ZZ^2$-periodic graph
$\Lambda$.} \label{fig:symmetric}
\end{center}
\end{figure}

Figure \ref{fig:symmetric} depicts the fundamental domain of a
more symmetric periodic graph.

We now define the Floquet Laplacian $\Delta( k ):\ell^2(V)\to
\ell^2(V)$ as the restriction of $\Delta$ defined as in
(\ref{eq:uno}) to the space $\ell^2_{ k }$. In terms of the
basis of the delta functions at vertices of $V$, this operator has
the following matrix:
\begin{equation}
\label{eq:cinque:sym} \Delta( k )\coloneq \left(
\begin{array}{cc}
A & B \\
B^{\dag} & A
\end{array} \right),
\end{equation}
where $A$ is the matrix
\begin{equation}
  \label{eq:cinque:A:sym}
  A( k )\coloneq  \left(
\begin{array}{ccccc}
0 & 1 & \rme^{-\rmi k_1} & 1 & 0 \\
1 & 0 & 1 & 0 & 0 \\
\rme^{\rmi k_1} & 1 & 0 & 1 & 1 \\
1 & 0 & 1 & 0 & \rme^{\rmi k_1} \\
0  & 0 & 1 & \rme^{-\rmi k_1} & 0
\end{array}\right),
\end{equation}
and $B$ is
\begin{equation}
  \label{eq:cinque:B:sym}
  B( k )\coloneq  \left(
\begin{array}{ccccc}
1 & 0 & 0 & 0 & 0 \\
0 & 0 & 0 & 0 & 0 \\
0 & 0 & \rme^{\rmi k_2} & 0 & 0 \\
0 & 0 & 0 & \rme^{\rmi k_2} & 0 \\
0 & 0 & 0 & 0 & 1
\end{array}\right).
\end{equation}
The matrix $B$ describes the interaction between the two symmetric
halves of the fundamental domain $V$.

In a similar way, one can define $L( k )$ and observe that
\begin{equation}
  \label{eq:sei:sym}
  L( k )= \left(
\begin{array}{cc}
T^{-1}& 0\\ 0 & T^{-1}
\end{array} \right)  \Delta ( k )
\left(
\begin{array}{cc}
T^{-1}& 0\\ 0 & T^{-1}
\end{array} \right),
\end{equation}
where $T$ is the matrix
\begin{equation}
  T\coloneq \left(
\begin{array}{ccccc}
2 & 0 & 0 & 0 & 0 \\
0 & \sqrt{2} & 0 & 0 & 0 \\
0 & 0 & \sqrt{5} & 0 & 0 \\
0 & 0 & 0 & 2 & 0 \\
0 & 0 & 0 & 0 & \sqrt{3} \\
\end{array}
\right).
\end{equation}

The fundamental domain has 10 vertices, and so there are 10 bands
to the spectrum. The lowest 5 bands are approximately
$[-3.840,-2.265]$, $[-2.943,-1.834]$ $[-1.865,-1.113]$,
$[-1.536,-0.333]$ and $[-0.803,0.377]$. The spectrum is symmetric
about $\lambda=0$, and the remaining five bands are reflections of
the previously mentioned bands about this point.

\begin{figure}[ht]
\begin{center}
\setlength{\unitlength}{4cm}
\begin{picture}(2,1.2)
\put(0.05,0.05){\includegraphics[angle=0,width=4cm,height=4cm]{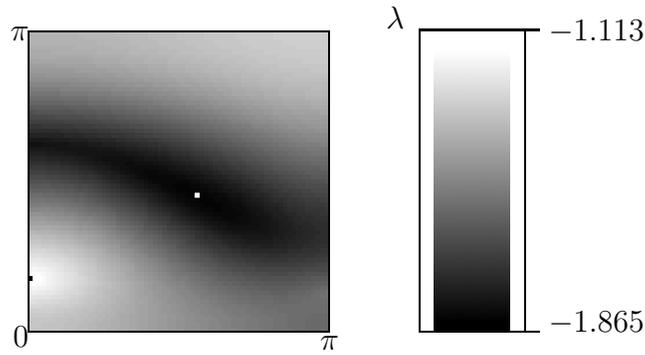}}
\put(0.05,0.05){\line(1,0){1}} \put(1.05,0.05){\line(0,1){1}}
\put(1.05,1.05){\line(-1,0){1}} \put(0.05,1.05){\line(0,-1){1}}
\put(1.4,0.05){\includegraphics[angle=0,width=1cm,height=4cm]{greyscale.eps}}
\put(1.35,0.05){\line(1,0){0.4}} \put(1.35,1.055){\line(1,0){0.4}}
\put(1.35,0.05){\line(0,1){1.005}}
\put(1.7,0.05){\line(0,1){1.005}} \put(1.24,1.06){$\lambda$}
\put(1.78,1.02){$-1.113$} \put(1.78,0.05){$-1.865$} \put(0,0){$0$}
\put(-0.01,1.02){$\pi$} \put(1.02,-0.01){$\pi$}
\end{picture}
\caption{A gray scale image of the third branch of the spectrum of
$\Delta ( k )$.
 Extreme points are highlighted.}
\label{fig:sym:sheetC}
\end{center}
\end{figure}

\begin{figure}[ht]
\begin{center}
\setlength{\unitlength}{4cm}
\begin{picture}(2,1.2)
\put(0.05,0.05){\includegraphics[angle=0,width=4cm,height=4cm]{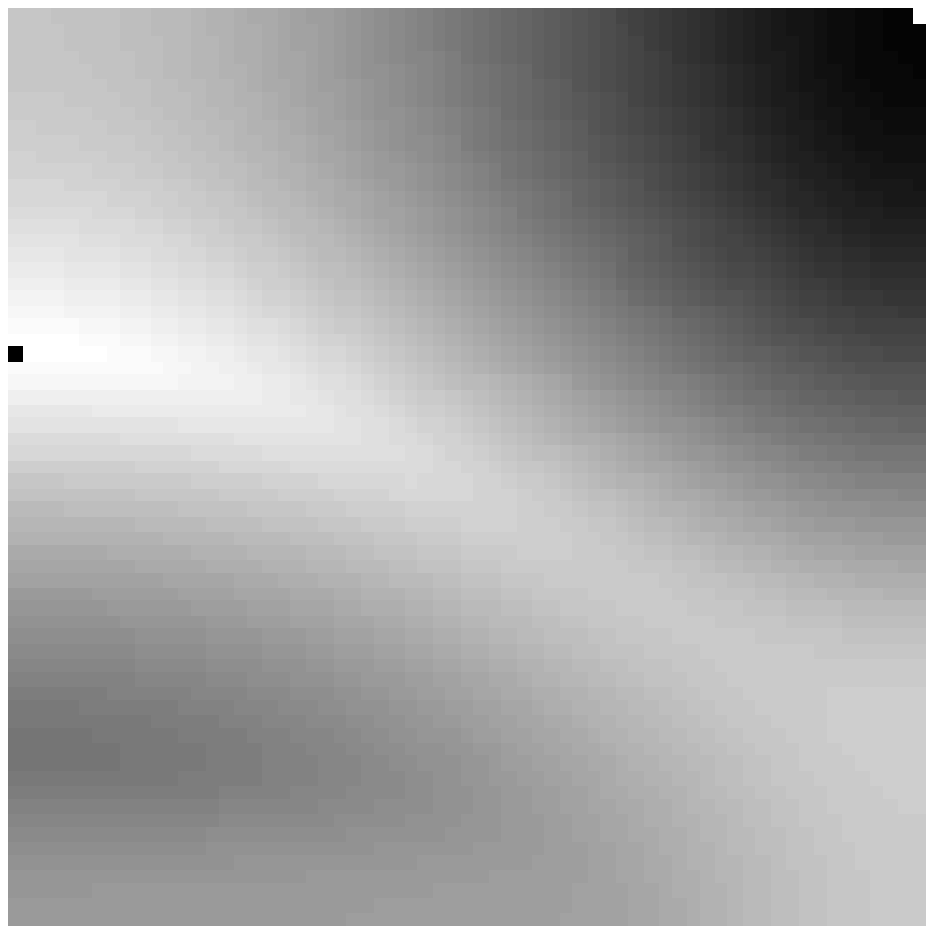}}
\put(0.05,0.05){\line(1,0){1}} \put(1.05,0.05){\line(0,1){1}}
\put(1.05,1.05){\line(-1,0){1}} \put(0.05,1.05){\line(0,-1){1}}
\put(1.4,0.05){\includegraphics[angle=0,width=1cm,height=4cm]{greyscale.eps}}
\put(1.35,0.05){\line(1,0){0.4}} \put(1.35,1.055){\line(1,0){0.4}}
\put(1.35,0.05){\line(0,1){1.005}}
\put(1.7,0.05){\line(0,1){1.005}} \put(1.24,1.06){$\lambda$}
\put(1.78,1.02){$-1.834$} \put(1.78,0.05){$-2.943$} \put(0,0){$0$}
\put(-0.01,1.02){$\pi$} \put(1.02,-0.01){$\pi$}
\end{picture}
\caption{A gray scale image of the second branch of the spectrum
of $\Delta ( k )$.
 Extreme points are highlighted.}
\label{fig:sym:sheetB}
\end{center}
\end{figure}

We again focus on a single example: the third band
$[-1.865,-1.113]$. A greyscale plot of the reduced Brillouin
zone\footnote{Note that for this figure, the {\em reduced}
Brillouin zone $[0,\pi]^2$ is plotted. The picture for the full
Brillouin zone is obtained by reflection of this picture in 2
directions.} for the solution curve corresponding to this band is
given in Figure \ref{fig:sym:sheetC}. On the other hand, Figure
\ref{fig:sym:sheetB} represents the band $[-2.943,-1.834]$, for
which maxima and minima occur both on the boundary $\Xi$ of the
reduced Brillouin zone (albeit, not on $X$). For the graph
$\Lambda$ we have observed that only the lower edge of the third
band, and upper edge of the eighth band are away from the boundary
of the reduced Brillouin zone. All other band edges do occur on
these lines of symmetry of the Brillouin zone.

\begin{figure}[ht]
\begin{center}
\setlength{\unitlength}{5cm}
\begin{picture}(2.5,2.0)
\put(0.5,0.0){\includegraphics[angle=0,width=10cm,height=10cm]{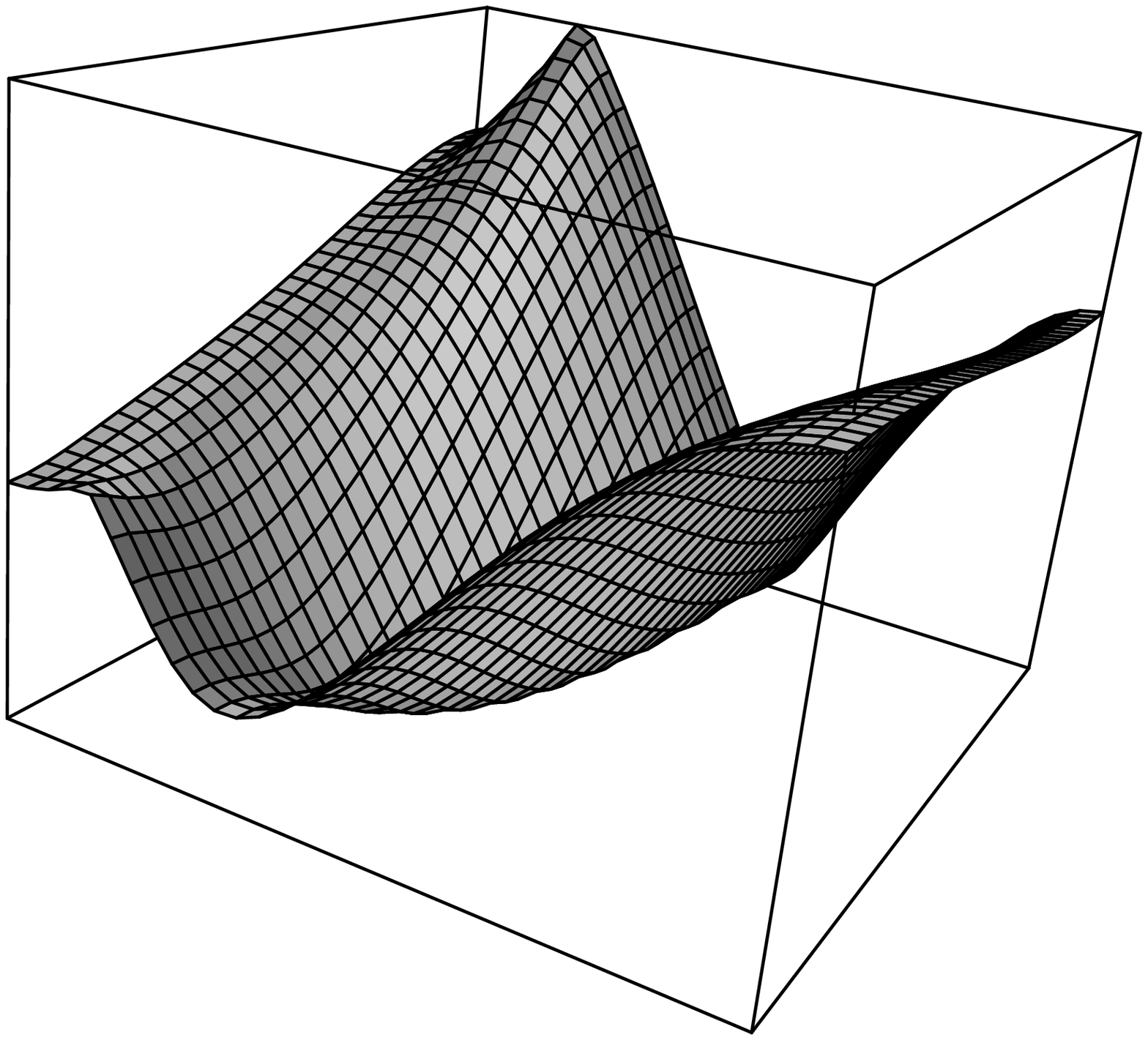}}
\end{picture}
\caption{A graph of the third branch.} \label{fig:sym_3d_sheetC}
\end{center}
\end{figure}

In order to check that the minimum point of the third branch
really occurs away from the boundary of the reduced Brillouin
zone, we repeat the procedure described in Section
\ref{SS:graphs}. With the aid of symbolic algebra computer
packages such as {\tt Maple} it is easy to find the characteristic
polynomial of the $10\times 10$ matrix \eqref{eq:cinque:sym}, and
to compute the derivative along the lines $k_1,k_2=0,\pi$. This
leads to a set of four pairs of polynomial equations (in the
variables $\lambda$ and $\cos(k_1)$ (or $\cos(k_2)$)) to be solved
simultaneously. Numerical root finding of this system reveals the
minimum along the boundary to occur at the point
$ k \approx(1.970,0)$ attaining a minimum value
$\lambda\approx -1.830$. We compare this with the strictly lower
point in the interior at $ k \approx(1.41,1.78)$ which takes
the value $\lambda\approx -1.865$, a difference of approximately
$2\%$ (or $5\%$ of the width of the band).

Looking only at the four ``corner'' points: $(0,0)$, $(0,\pi)$,
$(\pi,0)$, $(\pi,\pi)$, the minimum value attained on the third
band is $\lambda\approx-1.568$.

For the operator $L( k )$ the situation is qualitatively
similar. The positions of the various maxima and minima do not
move much. The minimum of the third band is located again away
from the boundary of the Brillouin zone. At the point
$ k \approx(1.28,1.93)$ the value attained by the third band
function is $\lambda\approx-0.5486$. It turns out that the minimum
value taken on the boundary of the reduced Brillouin zone is
$\lambda\approx-0.5380$ at $ k \approx(1.92,0)$, a difference
of $2\%$ (or $6\%$ of the width of the band).

One can also notice that we encounter here examples of spectral
edges occurring on $\Xi$, but not on $X$. So, all the
possibilities do materialize: spectral edges occurring on $X$
only, on $\Xi - X$, and finally on $B -\Xi$.

\section{Quantum graph case}\label{S:quantum}

In this section we consider the spectrum of a periodic quantum
graph $G$ with the same topology as $\Gamma$.  The spectrum
$\sigma (G)$ can be related to the spectrum of the Floquet
Laplacian $L(k)$ investigated in the previous section. As
a consequence, we will discover that the maxima and minima of some
branches, and thus spectral edges as well, occur at the same
quasi-momenta in both systems. Hence, spectral edges of this
periodic quantum graph Hamiltonian occur inside the Brillouin
zone.

We construct a metric graph $G$ by equipping all edges of $\Gamma$
with unit lengths. To complete the definition of a quantum graph,
we need to define a self-adjoint differential Hamiltonian. As
such, we consider the negative second derivative on the edges $e$
of $G$:

\beq\label{E:qhamilt} H=-\frac{\ud^2}{\ud x_e^2} \ . \eeq

The Hilbert space where the operator acts is $L^2(G)=\bigoplus _e
L^2(e)$. The domain of the operator consists of all functions $f$
such that
\begin{equation}\label{E:domain}
    \cases{
f\in H^{2}(e) \mbox{ for each edge } e,\\
\sum\limits_e \|f\|^2_{H^{2}(e)}<\infty,\\
f \mbox{ is continuous at each vertex}\\
\sum_{e\sim v} f'_e(v) = 0 \mbox{ (Kirchhoff, or Neumann,
conditions)}.}
\end{equation}
Here, $f'_e(v)$ denotes the outgoing derivative of $f$ at $v$
along the edge $e$.

It is well known (e.g., \cite{EG, Ku_graphs2, KuVainb, Oleinik})
and easy to check that Floquet theory applies to the quantum graph
case. In particular, the spectrum $\sigma(H)$ coincides with the
union over the Brillouin zone $B$ of the spectra of Floquet
Hamiltonians $H(k)$. Here $H(k)$ is the operator defined similarly
to $H$ on $H^2_{loc}$ functions with the same Kirchhoff vertex
conditions as in (\ref{E:domain}), and with the additional cyclic
(Bloch, Floquet) condition (\ref{E:cyclic}):
\begin{equation}
f(T( p  x))=\rme^{\rmi k \cdot  p  }f(x)
\end{equation}
for any $ p  \in\ZZ^2,\,x\in G$. We will call such functions
{\em Bloch (generalized) eigenfunctions}. Thus, describing the
spectrum of an either combinatorial, or quantum periodic graph
operator, we can work with such generalized eigenfunctions only.

 So, let $\psi$ be a Bloch eigenfunction of $H$ on $G$ with a
 quasimomentum $k$ and eigenvalue $\omega^2$,
\beq\label{eq:quantum2} H \psi = \omega^2 \psi \ . \eeq

Let us define a function $\varphi$ on the combinatorial
counterpart $\Gamma$ of $G$ as the restriction of $\psi$ to the
vertices of $G$. Clearly, $\varphi$ is a Bloch function with the
same quasimomentum $k$ (e.g., \cite{Ku_graphs2}). Due to
(\ref{eq:quantum2}), on each edge $e=(u,v)$ the function $\psi$
can be written in terms of its values $\varphi (v), \varphi(u)$ at
the endpoints:

\beq\label{eq:quantum3} \psi_e (x_e) = \varphi(v) \cos \omega x_e
+ \left( \frac{\varphi(u)-\cos \omega \, \varphi (v)  }{\sin
\omega}\right)
 \sin \omega x_e
\eeq

Here $\psi_e$ and $x_e$ denote the restriction of $\psi$ to the
edge $e$ and the coordinate along $e$ respectively. Then one has
\beq\label{eq:quantum4} \psi^\prime_e(v)=\frac{\varphi(u)-\cos
\omega \, \varphi(v)}{\tan \omega}. \eeq
Vertex conditions (\ref{E:domain}) imply that at each vertex $v$
the equation %
\beq\label{eq:quantum5} \frac{1}{d_v} \sum_{u\sim v} \varphi(u)=
\cos \omega \, \varphi(v) \ \eeq holds. Thus, $\varphi$ is a Bloch
eigenfunction of the difference operator $\frac{1}{d_v}
\sum_{u\sim v} \varphi(u)$ on $\Gamma$ with eigenvalue $\cos
\omega$.  Notice that the spectrum of this operator coincides with
the spectrum of its symmetrized version $L$ investigated in the
previous section. Thus, we have constructed a quantum graph
example of a periodic operator with a spectral edge attained
inside the Brillouin zone.

The reader might notice that the correspondence between the
spectra of $L$ and $H$ works smoothly only outside zeros
of $\sin \omega$, i.e. not on the Dirichlet spectrum of the edges.
This is a well known phenomenon, see for instance
\cite{Ku_graphs2}. However, it is easy to observe that this does
not influence our case and thus we have an example when a spectral
edge of a periodic quantum graph occurs inside the Brillouin zone.

\section{Neumann waveguides and periodic tubular manifolds}\label{S:guide}

In this section, we will show existence of periodic elliptic
second order operators on manifolds with a free co-compact action
of $\Z^2$, some of whose spectral edges are attained inside the
Brillouin zone. The simplest example is of the Laplace operator
with Neumann boundary conditions in a periodic planar waveguide.

In order to construct the guide, let us assume our graph $G$ (see
Figure \ref{fig:quantumfd}) to be embedded into the plane in such a
way
that:\\
1. Each edge is a smooth simple curve of length $1$.\\
2. Edges intersect only at the vertices.\\
3. Edges intersect transversally (i.e., there are no tangent edges).\\
4. The embedded graph is $\Z^2$-periodic.\\
Such an embedding is clearly possible.

Let us take a small $\epsilon >0$ and consider a ``fattened
graph'' domain $\Omega_\epsilon$ that consists of tubular
neighborhoods of the edges (domain $U$ in Figure \ref{Fig:fat} below) and
neighborhoods of vertices (domain $V$ in Figure \ref{Fig:fat}).

\begin{figure}[ht]
\begin{center}
\setlength{\unitlength}{5cm}
\begin{picture}(1.0,1.0)
\put(0.0,0.0){\includegraphics[angle=0,width=5cm,height=5cm]{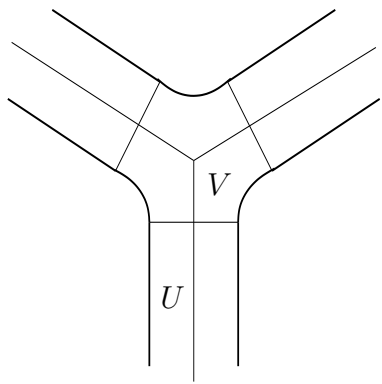}}
 \put(0.41,0.2){$U$}
 \put(0.535,0.5){$V$}
\end{picture}
\caption{A ``fattened graph'' domain.} \label{Fig:fat}
\end{center}
\end{figure}

We will assume the following conditions on the domain
$\Omega_\epsilon$:\\
1. The boundary is sufficiently smooth (e.g., $C^2$).\\
2. The domains $U$ have constant width $\epsilon$ in directions
normal to the edges.\\
3. The vertex neighborhoods $V$ satisfy the following property:
there exist balls $b_\epsilon$ and $B_\epsilon$ of radii
$r\epsilon$ and $R\epsilon$ correspondingly, centered at each
vertex and such that $b_\epsilon\subset V\subset B_\epsilon$.
Besides, $V$ must be star-shaped with respect to all points of
$b_\epsilon$.\\
4. The domain $\Omega_\epsilon$ is $\Z^2$-periodic.\\
It is easy to see that one can construct an $\epsilon$-dependent
family of domains satisfying all these properties.

Consider now the (positive) Laplace operator
$-\Delta_{N,\epsilon}$ in $\Omega_\epsilon$ with Neumann boundary
conditions on $\partial\Omega_\epsilon$.

It is proven in \cite{KZ,KZ2}, \cite{RS1}-\cite{RS4} that for any
value of the quasimomentum $k$ and any finite interval $I$ of the
spectral axis, the part in $I$ of the spectrum of the Floquet
operator $-\Delta_{N,\epsilon}(k)$ converges to the corresponding
part of the spectrum of the quantum graph Hamiltonian
(\ref{E:qhamilt})-(\ref{E:domain}) on $G$. Moreover, this
convergence is uniform with respect to $k$. This, in particular,
implies immediately the following

\begin{theorem}\label{T:guide}
For sufficiently small values of $\epsilon$, there is an isolated
band of the spectrum of the waveguide operator
$-\Delta_{N,\epsilon}$, whose end points are attained strictly
inside the first Brillouin zone.
\end{theorem}

\dimostrazione Indeed, this property holds for the quantum graph
Hamiltonian $H$ on $G$, and thus the convergence result shows that
it survives in $\Omega_\epsilon$ for small values of
$\epsilon$.\qed

This construction does not necessarily require the graph to be
planar. For instance, one would not be able to have a planar
embedding with required properties for the more symmetric graph
$\Lambda$ considered in Section \ref{SS:symmetric}. However, one
can embed $\Lambda$ into $\R^3$ in such
a way that it is $\Z^2$-periodic, with all other properties as
required before. Then a $3D$ waveguide domain $\Omega_\epsilon$
can be constructed around $\Lambda$ in a similar manner to the one
above, such that the statement of Theorem \ref{T:guide} still
holds.

\begin{figure}[ht]
\begin{center}
\scalebox{0.5}{\includegraphics{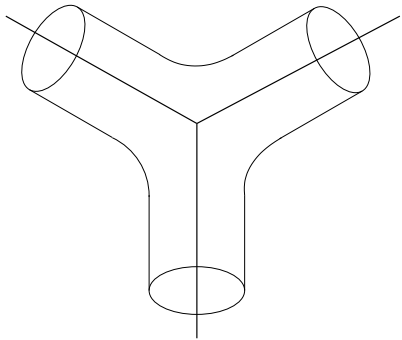}} \caption{A ``sleeve''
manifold.}\label{F:sleeve}
\end{center}
\end{figure}

Another type of examples can be constructed as a ``tight sleeve
Riemannian manifold'' $M_\epsilon$ around graphs $G$ or $\Lambda$.
The notion of such a manifold can be easily understood from the
Figure \ref{F:sleeve} (see precise definitions in \cite{ExnerPost}).

Such a manifold can be constructed preserving a co-compact free
action of $\Z^2$. Then the results of \cite{ExnerPost} concerning
convergence of spectra of the Laplace-Beltrami operator on
$M_\epsilon$ to those on the graph, show that the following result
holds:

\begin{theorem}\label{T:sleeve}
In any dimension $d\geq 2$, there exists an example of a closed
$d$-dimensional manifold $M$ with a co-compact, free, isometric
action of $\Z^2$, such that there is an isolated band of the
spectrum of the Laplace-Beltrami operator $-\Delta_M$, whose end
points are attained strictly inside the first Brillouin zone.
\end{theorem}
The proof coincides with the one of the previous theorem.

\section{Schr\"{o}dinger and Maxwell operators}\label{S:Schroed}

The previous discussion does not leave any doubt that examples can
be found for essentially any type of periodic equations of
mathematical physics. However, it is desirable to have such
explicitly described examples for the cases of periodic
Schr\"{o}dinger and Maxwell equations, interest in which stems
from the solid state and photonic crystal theories (e.g.,
\cite{AM}, \cite{FK1}-\cite{FK4}, \cite{JMW,JJ},
\cite{Ku}-\cite{Ku1}, \cite{RS}).

Although we do not currently have rigorous arguments to show the
existence of such examples, we can expect that they may be
obtained as follows. Consider a planar embedding of the graph $G$,
as the one considered in the previous section, with the additional
requirement that at each vertex the tangent lines to the
converging edges form equal angles. This can obviously be
achieved. Consider then a ``fattened graph'' domain
$\Omega_\epsilon$ described before and the Schr\"{o}dinger
operator $S\coloneq-\Delta+V(x)$ in $\RR^2$ with the $\Z^2$-periodic
potential $V(x)$ that is equal to zero in $\Omega_\epsilon$ and
equal to a large constant $C$ outside.

\begin{conjecture}\label{Con:Schroedinger}
Under appropriate asymptotics $\epsilon\rightarrow 0, C\rightarrow
\infty$, the spectrum of the operator $S$ will display an isolated
spectral band with its edges attained inside the Brillouin zone.
\end{conjecture}

What is lacking here, in spite of significant attention paid to
such asymptotics (e.g., \cite{Tenuta, DE2, ES, ExnerPost, ExS1,
Fr, FW, FH, Ku_CONM, Ku1, Ku_thin, KK1, KK2, KZ, KZ2, MolVai,
MolVai2, Post}, \cite{RS1}-\cite{Sa2}, \cite{Z}), is a spectral
convergence result analogous to the one for the Neumann Laplace
operator. Moreover, it is known that such convergence (even after
appropriate spectral re-scaling), does not hold, due to the appearance
of low energy (below the energy of the first transversal
eigenfunction) bound states attached to vertices
\cite{DE2,ExS1,Ku_thin}. However, for creating an example that we
are looking for, the full spectral convergence is not truly
needed. What is required, is some kind of convergence above the
energy of the first transversal eigenfunction, which must hold
(see some results in this direction in \cite{MolVai,MolVai2}).
When the angles formed by the edges are equal, we expect vertex
conditions of Kirchhoff type to arise.

Concerning the periodic Maxwell operators $\nabla \times
\epsilon^{-1}(x)\nabla\times$, where $\epsilon (x)$ is the
electric permeability, we expect that the simplest to come by will
be an example of a $2D$ periodic medium (i.e., a medium which is
periodic in two directions and homogeneous in the third one). As
it is well known \cite{Ja,JMW,JJ,Ku1}, in this case the Maxwell
operator splits (according to two polarizations) into the direct
sum of two scalar operators $-\nabla\cdot \epsilon^{-1}(x) \nabla$
and $-\epsilon^{-1}(x) \Delta$. We expect that similar
high-contrast narrow media as described above should provide
necessary examples (see also considerations of such high contrast
limits in \cite{AKK}, \cite{FK1}-\cite{FK4},\cite{Fr_dens,
HempLie, Ku1, KK1, KK2, Selden}).

\section{Why do spectral edges often appear
at the symmetric points of the Brillouin zone?}\label{S:positive}

In this section, we will show why, in spite of the examples of this
paper, the spectral edges are  often attained at the highest
symmetry points, and hence at the boundary of the reduced Brillouin
zone. Namely, let $H_0$ be a periodic self-adjoint elliptic
Hamiltonian with real coefficients (i.e., the corresponding
non-stationary Schr\"{o}dinger equation has time-reversal symmetry).
Suppose that the spectral edges of $H_0$ are attained at symmetry
points of the Brillouin zone $B=[-\pi,\pi]^n$ only (see the details
below). Then we will show that for a ``generic'' $H_0$, this feature
of the spectrum cannot be destroyed by small perturbations (with the
same symmetry) of the operator. This robustness might be the reason
why one very rarely observes spectral edges appearing inside the
reduced Brillouin zone.

Let us now introduce some notions.
We will assume, for simplicity of presentation,
that the Hamiltonian is the Schr\"{o}dinger operator
in $\RR^n$:
\begin{equation}\label{E:unpert}
H_0=-\Delta+ V(x),
\end{equation}
where $V(x)$ is a real-valued bounded potential such that
$V(x+p)=V(x)$ for all integer vectors $p\in\ZZ^n$.
For any quasimomentum $k\in B$, we will denote by
$H(k)$ the Bloch Hamiltonian defined on $\ZZ^n$-periodic
functions (i.e., on functions on the torus $\TT^n=\RR^n / \ZZ^n$)
as
\begin{equation*}
H (k)=\left(\frac{1}{\rmi}\nabla+k\right)^2+ V(x).
\end{equation*}
It depends polynomially, and thus analytically, on $k$. We denote
by $\lambda_j(k), j= 1, 2, \dots,$ the eigenvalues of $H(k)$
counted with their multiplicity in non-decreasing order. The band
functions  $\lambda_j(\ \cdot\ )$  are continuous functions of
$k\in B$. Since the potential $V$ is real-valued, the eigenvalues
are also even in $k$, i.e. $\lambda_j(-k) = \lambda_j(k)$. This
follows from the fact that complex conjugate to an eigenfunction
is an eigenfunction (presence of a magnetic potential would
destroy this symmetry). This symmetry will be crucial for what
follows.

The ranges
\begin{equation*}
\Delta_j=\{\lambda_j(k)|\,k\in B\}
\end{equation*}
are closed finite intervals of the spectral axis ({\em spectral
bands}), whose union is the spectrum $\sigma(H_0) $. Global maxima
and minima of the band functions $\lambda_j(\ \cdot\ )$ are the endpoints (edges) of
spectral bands.
It is also known (e.g., \cite{Kato, Ku,RS}) that $\lambda_j$'s are
analytic in $k$ away from the eigenvalue crossing points.
In case of the crossing,
we point out the following elementary, but useful result:

\begin{lemma}\label{l:crossing}
Let us fix an open interval $\Delta=(a,b)\subset \R$. Suppose that
for a neighborhood $U\subset B$ the band functions $\lambda_s$,
$j\le s\le j+m$, satisfy
\begin{equation*}
\lambda_s(k)\in \Delta,\ k\in U,
\end{equation*}
and that the remaining band functions take values in $U$ that lie outside
a neighborhood of the closed interval $\overline{\Delta}$. Then the functions
\begin{equation}\label{E:det_tr}
\prod_{s=j}^{j+m} \lambda_s(k) \mbox{ and }
\sum_{s=j}^{j+m} \lambda_s(k)
\end{equation}
are analytic with respect to $k\in U$.
\end{lemma}

\begin{proof} Let us assume that $m= 1$ (the case of arbitrary $m$ works out
exactly same way), i.e.\ we have two eigenvalue branches
$\lambda_-(k) \coloneq \lambda_j(k)$ and
$\lambda_+(k)\coloneq\lambda_{j+1}(k)$. Consider a positively
oriented circle $\Gamma\subset \CC$ centered at $(a+b)/2$ with
radius $(b-a)/2+\epsilon$ with a small $\epsilon >0$. The
two-dimensional projection
\begin{equation}\label{E:projection}
P(k)=\frac{1}{2\pi \rmi}\int\limits_\Gamma
\left(\lambda-H(k)\right)^{-1}d\lambda,
\end{equation}
depends analytically on $k\in U$. Let $M(k)$ be the range
of $P(k)$. It forms an analytic two-dimensional vector-bundle
\cite{Ku,ZK}. Let $\{e_1,e_2\}$ be a basis in $M(k_0)$ with some $k_0\in U$.
For $k$ close to $k_0$ we can define a basis of $M(k)$
analytically depending on $k$ as follows:
\begin{equation}\label{E:basis}
f_j(k)\coloneq P(k)e_j.
\end{equation}
In this basis, the operator function
\begin{equation*}
P(k)H(k)P(k)|_{M(k)}=H(k)P(k)|_{M(k)}
\end{equation*}
can be written as an analytic $2\times 2$ matrix-function $A(k)$
with eigenvalue branches $\lambda_-(k)$ and $\lambda_+(k)$.
Therefore, the functions
\begin{equation*}
\det A(k), \ \ \textup{tr}\ A(k)
\end{equation*}
are analytic in $k$ in a neighborhood of $k_0$.
\end{proof}

The only  fixed points $k\in B$
for the symmetries $k\to-k+p,p\in 2\pi\Z^n$, are the ones from the set
\begin{equation}\label{E:xi}
X=\{k=(k_1,\cdots,k_n)\in B\,|\, k_j\in\{0, \pi\},j=1,\cdots,n\}.
\end{equation}
In view of the symmetry $\lambda_j(k) = \lambda_j(-k)$ we also have
$\lambda_j(k_0+ k) = \lambda_j(k_0-k)$ for any $k_0\in X$. We have
already shown in this text that the global extrema of the band
functions can occur outside the set $X$. The experimental
observation, however, is that for most periodic operators of
practical importance and for practical values of their parameters
(e.g., potentials, electric permittivity, etc.), the band endpoints
do occur on $X$. One can easily observe this by looking at
dispersion curve calculations in solid state physics or photonic
crystals literature (e.g., \cite{Bibl, Busch}). {\bf Our main question now
is: Why do the spectral edges occur so often on $X$?}

Our considerations will be local on the spectrum. Thus, let us fix
a finite interval $\Lambda=(a,b)$ of the spectral axis.
Note that the number of spectral bands $\Delta_j$ overlapping with
$\Lambda$ is finite.
We first introduce the following notion:

\begin{definition}\label{d:simple}
We call a periodic Hamiltonian $H$ {\bf simple} on a
finite interval $\Lambda$,
if the global extrema of the band functions $\lambda_j(k)$ which occur inside $\Lambda$,
are attained at the points of the set $X$ only.
\end{definition}

The simplicity property defined above will be discussed for
``generic'' periodic operators:

\begin{definition}\label{d:generic}
We call a periodic Hamiltonian $H$ {\bf generic} on a finite interval
$\Lambda$,
if for every band edge $\lambda_0$ occuring inside $\Lambda$,
the band functions $\lambda_j$ assume the value $\lambda_0$ at finitely many
points of the Brillouin zone $B$, and in a neighborhood $U$ of each such point
$k_0$
one of the following two conditions is satisfied:
\begin{enumerate}
\item
There is a unique band function $\lambda(\ \cdot \ )$ for $k\in U$
such that $\lambda(k_0) = \lambda_0$; moreover, $k_0$ is a
non-degenerate extremum of $\lambda(k)$.
\item
For $k\in U$ there are only two band functions
$\lambda_+(\ \cdot\ )$, $\lambda_-(\cdot\ )$ such that
$\lambda_-(k_0)= \lambda_+(k_0) = \lambda_0$. Moreover,
$\lambda_-(k)< \lambda_+(k)$ for all $k\in U - \{ k_0\}$, and
$k_0$ is
a non-degenerate maximum of the product
$D(k) = (\lambda_+(k) - \lambda_0)(\lambda_-(k) - \lambda_0)$.
\end{enumerate}
Above by a non-degenerate extremum  we
understand an extremum with a non-degenerate Hessian.
\end{definition}

Recall that in view of Lemma \ref{l:crossing}, the determinant of
$A(k)$ is analytic on $U$. Definition \ref{d:generic} means that the
band functions of a generic Hamiltonian behave near band edges as
eigenvalues of a ``generic'' $2\times 2$ self-adjoint analytic
matrix function. We refer to cases (1) and (2) in the above
definition as \textit{the single edge case} and \textit{the case of
two touching bands} respectively. Note also that in Definition
\ref{d:generic} the band edge $\lambda_0$ is not assumed to be
(albeit could be) an endpoint of the spectrum.

The following conjecture is believed to hold (see a variety of similar
genericity conjectures in, e.g.,
\cite{Avron3,Colin2, Ku1,Novikov_VINITI}).

\begin{conjecture}\label{conjecture}
Generic periodic Hamiltonians form a set of second Baire category
in a suitable class of periodic operators.
\end{conjecture}

The closest to the proof of this conjecture is the
result of \cite{Klopp}, where it was shown that generically a band
edge, which is an endpoint of the spectrum,
is attained by a single band function.

Our aim is to show
that a generic simple Hamiltonian $H_0$
remains generic and simple under small perturbations. More precisely, we introduce
the family of operators
\begin{equation*}
H_g=H_0+gV(x), \ H_0 = -\Delta + V_0,
\end{equation*}
where $V_0$ and $V$ are bounded real-valued $\Z^n$-periodic functions, and
$g\in \R$ is a parameter. We denote by $\lambda_j(k, g)$ the band functions
of $H_g$. If $g = 0$, we drop $g$ and write simply $\lambda_j(k)$. Since $H_g$
is analytic in $g$, the band functions $\lambda_j(k, g)$ are analytic in
$(k, g)$ away from the crossing points, and the quantities
defined in \eqref{E:det_tr}
are analytic in $(k, g)$ under the conditions of Lemma \ref{l:crossing}.

We can now formulate a result that gives a partial answer to the
question posed in this Section.

\begin{theorem}\label{T:condit}
Let $\Lambda \subset\R$ be a finite closed interval, and let the
operator $H_0$ (see \eqref{E:unpert}) be simple and generic in a
neighborhood of $\Lambda$. Then, for sufficiently small values of
$g$, the operator $H_g$ is also simple and generic in a
neighborhood of $\Lambda$.
\end{theorem}

\dimostrazione Let $\Lambda'$ be a finite closed interval
containing $\Lambda$ in its interior and such that operator $H_0$
is simple and generic in an open neighborhood $\Lambda''$ of
$\Lambda'$. The continuity of $\lambda_j(k, g)$ in $g$ guarantees
that for small $g$, the spectral band edges of the perturbed
operator occurring on $\Lambda'$, are either perturbations of the
band edges of $H_0$ that are inside $\Lambda''$, or are produced
by opening a gap between two touching spectral bands of $H_0$.

Let $\lambda_0\in \Lambda''$ be a single band edge of $H_0$, or
the point where two bands touch. Assume without loss of generality
that $\lambda_0 = 0$. Since the unperturbed operator $H_0$ is
simple, again, by continuity of $\lambda_j(k, g)$ in $g$, for
sufficiently small values of $g$, the perturbed eigenvalues cannot
reach their global maxima outside a neighborhood of the set $X$.
Thus, it suffices to consider the neighborhood of each point
$k_0\in X$ individually. We assume without loss of generality that
$k_0 = 0$.

Further proof requires different arguments for the
two cases featuring in Definition \ref{d:generic}.

\subsection{The single edge case}\label{SS:deformation}
Let $\lambda(k)$ be the unique band function of the operator $H_0$,
which attains at $k_0 = 0$ its
non-degenerate extremum, which for definiteness will be assumed
to be a maximum. Recall that $\lambda(\ \cdot\ )$ is analytic
in $k$ and $\lambda(k) =  \lambda(-k)$, so that
\begin{equation*}
\lambda(k) = \lambda_2(k) + \lambda_e(k),
\end{equation*}
where $\lambda_2$ is a negative definite quadratic form and
$\lambda_e$ is an analytic function such that
$\lambda_e(k)=\Or(|k|^4)$. For sufficiently small $g$ and $k$, the
eigenvalue $\lambda(k, g)$ will remain separated from the rest of
the spectrum of $H_g(k)$. Thus, to complete the proof, we need to
show that $\lambda(\ \cdot\ , g)$ attains its maximal value at $k =
0$ and this maximum is non-degenerate. Due to analyticity,
\begin{equation*}
\lambda(k,g)=\lambda_2(k)+\lambda_e(k)+g\widetilde{\lambda}(k,g),
\end{equation*}
where $\widetilde{\lambda}(k,g)$ is a real-valued real-analytic function
of $(k, g)$, and $\widetilde{\lambda}(k,g) = \widetilde{\lambda}(-k,g)$.
The latter property implies that
\begin{equation*}
\nabla_k\widetilde{\lambda}(k,g) = \Or(|k|),
\end{equation*}
uniformly in $g$.
Making an appropriate linear change of variables, we can always assume that
$\lambda_2(k) = -|k|^2/2$. Then, taking the
gradient with respect to $k$, we obtain
\begin{equation*}
\nabla_k \lambda(k,g)=-k+\nabla_k\lambda_e(k)+g\nabla_k\widetilde{\lambda}(k,g).
\end{equation*}
Consequently
\begin{equation*}
|\nabla_k\lambda_e(k)+g\nabla_k\widetilde{\lambda}(k,g)|\leq C(|k|^3 + g|k|),
\end{equation*}
with some positive constant $C$, and hence,
for $|g|<(4C)^{-1},|k|<(2\sqrt{C})^{-1}, k\neq0,$ we get,
\begin{equation*}
|\nabla_k \lambda(k,g)|>\frac{|k|}{2}\neq 0.
\end{equation*}
This proves that the only stationary point of $\lambda(\ \cdot\ , g)$ is $k=0$.
Moreover, since $\lambda_2(\ \cdot\ )$
is negative definite, the function $\lambda(\ \cdot\ , g)$ has a non-degenerate
Hessian if $g$ is sufficiently small. Thus the band function
$\lambda(k, g)$ attains its extremum on $X$ and satisfies the requirements
of Definition \ref{d:generic}(1).

\subsection{The case of two touching bands}
Assume, as above, that $\lambda_0 = 0$, $k_0 = 0$, and that
$\lambda_-(k)$ and $\lambda_+(k)$ are the band functions as given in
Definition \ref{d:generic}. Denote by $\lambda_{\pm}(k, g)$ the perturbed
band functions.
According to Lemma \ref{l:crossing}, the functions
\begin{equation*}
d(k, g) = \lambda_-(k, g)\lambda_+(k, g),\
t(k, g) = \frac{1}{2}\bigl(\lambda_-(k, g)+\lambda_+(k, g)\bigr)
\end{equation*}
are analytic in a neighborhood of $(k, g) = (0, 0)$.
Remembering the central symmetry of the eigenvalues and the genericity assumption
for $H_0$, we can write
\begin{equation}
\eqalign{
d(k,g) = d_2(k)+d_e(k)+g\widehat{d}(k,g),\\
t(k,g)=t_2(k)+t_e(k)+g\widehat{t}(k,g).}\label{E:det_tr_repres}
\end{equation}
Here all functions are analytic near $(k, g) = (0,0)$ and even in
$k$. The functions $t_2$ and $d_2$ are quadratic forms,
the terms $d_e,t_e$ are $\Or(|k|^4)$, and, by virtue of genericity,
$d_2$ is negative definite. Thus, as in the first part of the proof,
we may assume that $d_2(k) = -|k|^2/2$.
Note also,  that $\hat d(0, g) = 0$, since
the eigenvalues $\lambda_{\pm}(0, g)$ are of order $\Or(g)$, and hence
$d(0, g) = \Or(g^2)$.
Introduce the quantity
\begin{equation*}
m = t^2 - d = \frac{1}{2}(\lambda_+ - \lambda_-)^2\ge 0.
\end{equation*}
Using \eqref{E:det_tr_repres} we get
\begin{equation*}
m(k, g) = g^2 m_0(g) - d_2(k) + m_e(k, g),\ m_e(k, g) = \Or(|k|^2) (k^2 + g),
\end{equation*}
where the functions
$m_0(g) = \hat t(0, g)^2 - \partial_g \hat d(0, g)$  and
$m_e(k, g)$ are analytic in $k, g$, and $m_e$ is even in $k$. Since
$m(0, g) = g^2 m_0(g)$, we also have $m_0(g)\ge 0$.

Let us list some simple estimates that these functions and their
gradients with respect to $k$ satisfy in a neighborhood of $(0,0)$. Below
we denote by $C, C_1$ some positive constants whose precise value is not important:
\begin{equation}
\eqalign{
|t(k,g)|\leq C(k^2 + g),\ |m(k,g)|\leq C(|k|^2+g^2),\\
|\nabla_k t(k,g)| , |\nabla_k m(k,g)|\leq C|k|,}\label{E:aux_dt}
\end{equation}

\begin{equation}
\eqalign{
|\nabla_k m(k, g)|\geq \frac{1}{2}|k|,\\
m(k, g)\geq g^2 m_0(g) + \frac{1}{4}|k|^2.}\label{E:nondeg}
\end{equation}
The eigenvalues  $\lambda_{\pm}(k, g)$ solve the characteristic
equation
\begin{equation*}
\lambda^2-2t(k,g)\lambda +d(k,g)=0,
\end{equation*}
and thus
\begin{equation}\label{E:eig}
\lambda_{\pm}(k, g)=t(k,g)\pm \sqrt{m(k,g)}.
\end{equation}
By \eqref{E:nondeg} the eigenvalues $\lambda_{\pm}(k, g)$ can coincide only at
$k = 0$, so that they are analytic in $k$ for $k\not = 0$. Let us prove that
$\lambda_{\pm}(\ \cdot\ , g)$  have no stationary points if $k\not = 0$.
Differentiate:
\begin{equation*}
\nabla_k \lambda_{\pm}(k, g)=\nabla_k t(k,g)\pm \frac{ \nabla_k m(k, g)}{2\sqrt{m(k,g)}}.
\end{equation*}
Now the estimates \eqref{E:aux_dt} and \eqref{E:nondeg} imply:
\begin{equation*}
|\nabla_k \lambda_{\pm}|\geq \left|\frac{ \nabla_k m}
{2\sqrt{m}}\right|-|\nabla_k t|\geq \frac{c|k|}
{ \sqrt{|k|^2+g^2}}-C|k|\geq C_1|k|
\end{equation*}
for small $g$ and $k\neq 0$.
This proves that $\lambda_{\pm}(\ \cdot\ , g)$ attain their extrema only
at $k = 0$.

It remains to show that the eigenvalues satisfy the requirements either of
Part (1) or Part (2) of Definition \ref{d:generic}.
If $m_0(g)> 0$, then by \eqref{E:nondeg}
and \eqref{E:eig}, the eigenvalues
$\lambda_+(k, g)$ and $\lambda_-(k, g)$ are decoupled for all $k$ and $g$,
and their extrema  are clearly non-degenerate.
If $m_0(g)=0$, then by \eqref{E:eig}
$\lambda_+(0, g) = \lambda_-(0, g)$, and then by \eqref{E:det_tr_repres} the determinant
$d(k, g)$ has a non-degenerate Hessian for small  $k, g$.

The proof of the \ref{T:condit} is complete.
\qed

\section{Final remarks}\label{S:remarks}
\begin{itemize}
\item Suppose that for a particular periodic operator the spectral
edges do occur at the point of $X=\{k|\, k_j=\pm \pi \mbox{ or }
0\}$ only. This means then that one can find the correct spectral edges
(and thus the spectrum as a set), computing only spectra of
problems that are periodic or anti-periodic with respect to each
variable (say, periodic with respect to $x_1$ and $x_3$ and
anti-periodic with respect to $x_2$). This resembles then the $1D$
situation \cite{Ea,RS}, when the edges of the spectrum are
attained at the spectra of the periodic and anti-periodic
problems.

\item In the last Section we have restricted ourselves  to the case of
Schr\"{o}dinger operators with electric potentials only. However,
the proof in fact does not use the structure of the operator and
could be extended to arbitrary analytically fibered operator in the sense
of \cite{Gerard}, as long as the central symmetry $k\mapsto -k$ holds.

We also assumed parametric perturbation (i.e., perturbation by $gV$
with a small scalar parameter $g$). However, without any change in
the proof, one can consider $V$ as a functional parameter and prove
the same statements for small values of this parameter.

\item Observations of stability under small perturbations
of critical points of a function in ``general position'' in presence
of symmetries, analogous to the ones in the last section, have been
made before in different circumstances, e.g. in \cite{Helffer}.
\end{itemize}

\section*{Acknowledgements}

The authors thank G.~Berkolaiko, K.~Busch, D.~Dobson, B.~Helffer,
A.~Tip, M.~Weinstein, and Ya~Yan~Lu for discussing the topic of this
paper.

The research of the second author was partly sponsored by the NSF
through the Grant DMS-0406022. The work of the first and fourth
authors was partly sponsored by the NSF Grant  DMS-0604859. The
authors thank the NSF for this support.

Part of the work by the second and third authors was completed
during the Isaac Newton Institute program on Spectral Theory in
July 2006. The authors thank the INI for the support.

\end{document}